\documentclass{article}
\usepackage{graphicx}
\usepackage{url}
  \let\oldurl\url
\usepackage{hyperref}
  \let\linkurl\url
  \let\url\oldurl
\usepackage[english]{babel}

\title{
\includegraphics[width=0.35\textwidth]{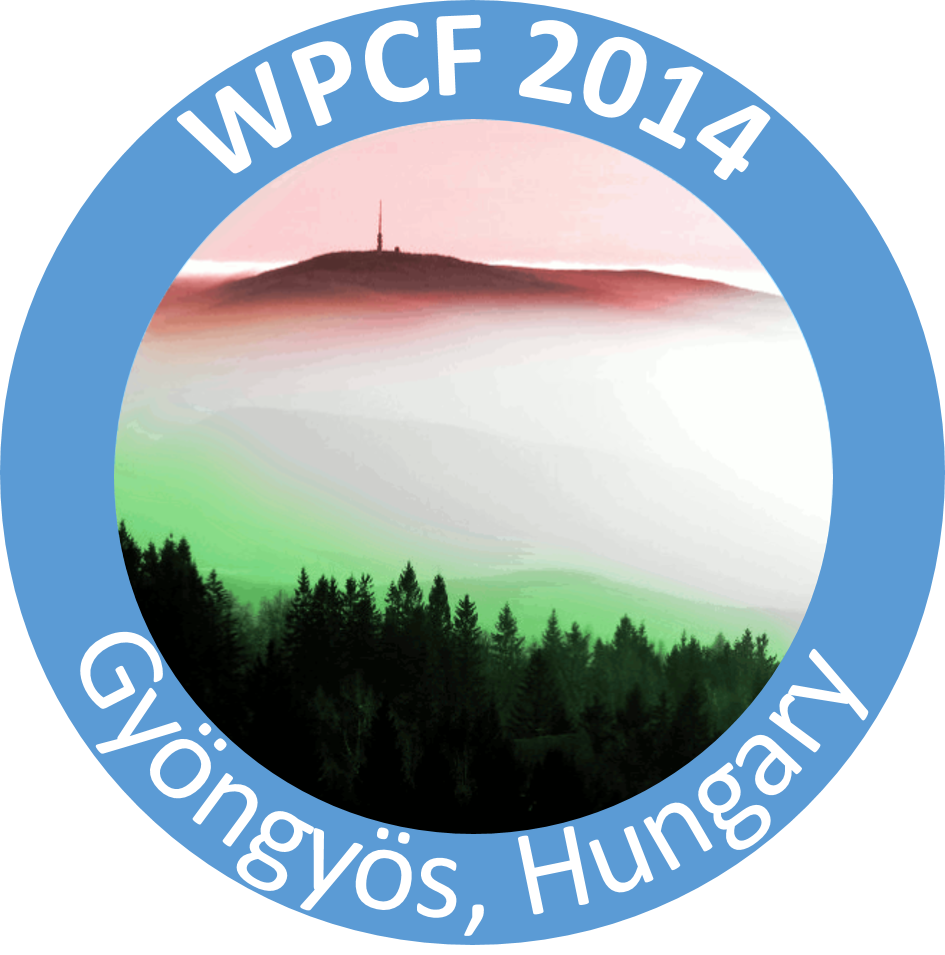}\\[1cm]
Qbe: Quark Matter on Rubik's Cube
\footnote{
\fontfamily{lmss}\selectfont
Dedicated to the 10th anniversary of the discovery of the perfect fluid of
quarks at RHIC as well as to the 40th anniversary of the invention of Rubik's
Cube.
}
}
\author{{T. Cs\"org\H{o}$^{1,2}$}\\[1ex]
$^1$EKE KRC, H-3200 Gy{\"o}ngy{\"o}s, M\'atrai u. 36, Hungary\\
$^2$Wigner RCP, H-1121 Budapest XII, Konkoly-Thege 29-33, Hungary
}

\begin{document}

\fontfamily{lmss}\selectfont
\maketitle

\begin{abstract} 
Quarks can be represented on the faces of the 3x3 Rubik's cube
with the help of a symbolic representation of quarks and anti-quarks, that was
delevoped originally for a deck of elementary particle cards, called Quark
Matter Card Game. Cubing the cards leads to a model of the nearly perfect fluid
of Quark Matter on Rubik's cube, or Qbe, which can be utilized to provide
hands-on experience with the high entropy density, overall color neutrality and
net baryon free, nearly perfect fluid nature of Quark Matter.  
\end{abstract}

\section{Introduction}

\begin{figure}[ht]
        \centering
        \includegraphics[width=0.52\linewidth]{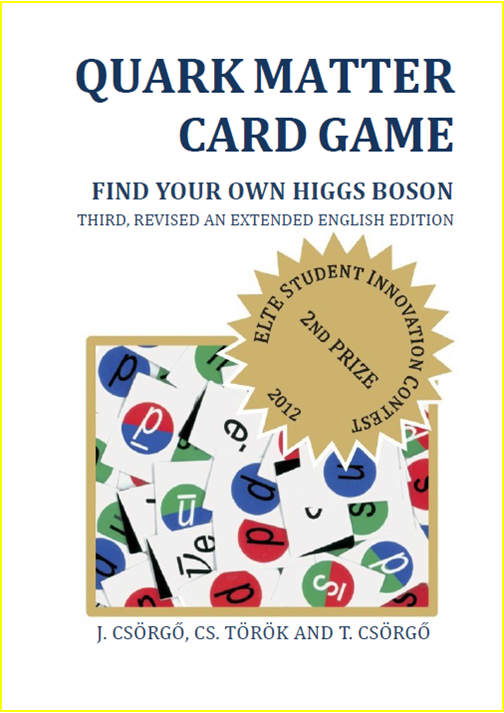}~\\*
        \caption{
\fontfamily{lmss}\selectfont
English language edition of Quark Matter Card Game
that describes games with a deck of elementary
particle cards, including a model the Early Universe just a
few microseconds after the Big Bang~\cite{Csorgo:2012qm}. 
Pick up a deck of  Quark Matter Cards and you can play 
heavy ion collisions  too,
for a tiny fraction of the cost of doing an actual experiment 
at the RHIC or LHC accelerators.
}
        \label{fig_1}
\end{figure}

In 2011, Cs. T\"or\"ok, a 17 years old secondary school student in studying
Gy\"ongy\"os, Hungary invented a card game with elementary
particles~\cite{Csorgo:2012qm}.  By 2014, this Quark Matter Card Game became an
invention, a patent and a product.  Initially, four different kind of games
were described in the first edition of a the book ``Quark Matter Card Games -
Elementary Particles, Playfully" playable with the same  deck of 66 cards,
representing elementary particles from the Standard Model of particle physics.
By now about a dozen of various card games are invented, all based on the same
deck of Quark Matter Card Game.  Some of these games are described in the
public domain, like the memory style quark matter card
game~\cite{Csorgo:2013xza} (where pairs or triplets of particle cards are to be
remembered) and its advanced version, called "Find your own Higgs boson"
~\cite{Csorgo:2013vza} where a Higgs boson is identified from its leptonic
decay modes, that requires to remember four cards in an advanced, memory style
game. 
In the so called Quark Matter Card Game, Figure~\ref{fig_1}, the players can
familiarize themselves not only with some of the elementary particles that are
the fundamental constituents of matter, but also with the properties of Quark
Matter, the recently discovered new phase of matter, that behaves not as a gas
but as a perfect fluid of quarks. Such a perfectly flowing Quark Matter filled
our Early Universe just a few microseconds after the Big Bang.  In 2004, this 
old-new state of matter was discovered in high energy heavy ion collisions
at the RHIC accelerator at BNL, located on Long Island near New York, NY, US.
Subsequently, the properties of Quark Matter were confirmed at larger colliding
energies at the LHC accelerator, located beneath the France-Switzerland border
near Geneva, Switzerland.


\vfill\eject 
Recently, an outdoor game called  ``Quark Wars" was also developed and tested,
that utilizes the deck of  Quark Matter Cards. Quark Wars is modelled on the so
called ``Hungarian number wars" outdoor game, with notable influence of the
American epic space saga ``Star wars"~\cite{Csorgo:2016idk}.

In addition to being a contribution to the proceedings of the WPCF 2014
conference, this manuscript is also an extended and updated version of a
handout booklet, distributed by the Guests, Users and  Visitors  Center of
Brookhaven National Laboratory at the 2015 AGS and RHIC Users Meeting, that was
dedicated to the 10th anniversary of the publications of the so called RHIC
White Papers, announcing the discovery of the prefect fluid of quarks
~\cite{Arsene:2004fa,Adcox:2004mh,Back:2004je,Adams:2005dq}.

\section{Anniversaries}

In 2014, we celebrated several anniversaries:

\begin{itemize}

\item 1944, 70 years before: Ern\H{o} Rubik was born in Budapest, Hungary ~\cite{Rubik:70}. 

\item 1954, 60 years before: CERN, the European Laboratory for Particle and Nuclear Physics was founded~\cite{CERN:60}.

\item 1974, 40 years before: Mr. Rubik created the prototype of his cube ~\cite{Rubik:70}.

\item 2004, 10 years before: The perfect fluid of quarks was discovered in
gold-gold collisions at BNL's RHIC accelerator 
~\cite{Arsene:2004fa,Adcox:2004mh,Back:2004je,Adams:2005dq}

\end{itemize}

In the followings we present, how one can ``dress up" or decorate a 3x3 Rubik's
Cube with colored quarks and anti-quarks, using a symbolic notation of quarks
and anti-quarks, as developed for the Quark Matter Card Game.  This manner,
Rubik's Cube becomes Qbe, a model or a symbolic representation of Quark Matter
on Rubik's 3x3  Cube, corresponding to a special Cube dedicated to the
promotion or popularization of the properties of the Perfect Fluid of Quarks.

The Perfect Fluid of Quarks or  Quark Matter is the hottest known form of
matter ever made by humans, with temperatures reaching above  $5 \times
10^{12}$ Kelvin in heavy ion collisions at CERN LHC~\cite{Guinness:2017}.  Such
a prefect fluid of quarks has been detected in the debris of high energy heavy
ion collisions at BNL's RHIC accelerator and the results were confirmed at
larger initial colliding energies at CERN's Large Hadron Collider (LHC). The
perfectness of Quark Matter  corresponds to its flowing properties: the
natural, internal scale of dissipative motion called kinematic viscosity of
this fluid is found to have the lowest value from among the known, human-made
materials.

\begin{figure}[ht]
        \centering
        \includegraphics[width=0.6\linewidth]{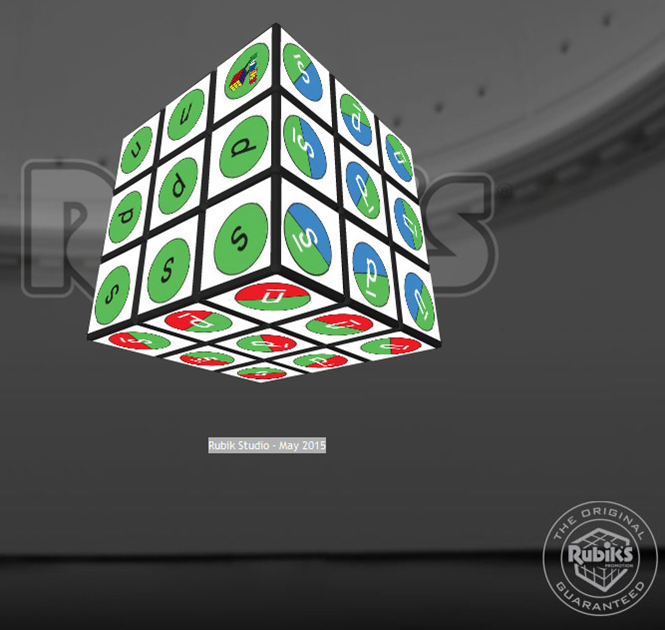}~\\*
        \caption{
\fontfamily{lmss}\selectfont
Qbe, the Quark Matter Cube,
representing the perfect fluid of quarks that filled the our Universe just a
few microsecond after the Big Bang. 
Image and the corresponding animation of Qbe is the courtesy 
of Rubik Studio Ltd.,~\cite{RubikStudio:2014an}.
}
        \label{fig_2}
\end{figure}

This conference contribution was first presented in 2014, at the 10th Workshop
on Particle Correlations and Femtoscopy. By now, quite some time has been passed 
since 2014, but at that time it was natural to dedicate the Quark Matter Cube
(in short, Qbe) to the 10th anniversary of the Perfect Fluid of quarks created
in gold-gold collisions at BNL's Relativistic Heavy Ion Collider.  The
artist's view of  Qbe is presented  on Figure ~\ref{fig_2} . 

\vfill\eject

\section{
Quark Matter on Rubik's Cube - Playfully
}

\begin{figure}[htb]
        \centering
        \includegraphics[width=0.9\linewidth]{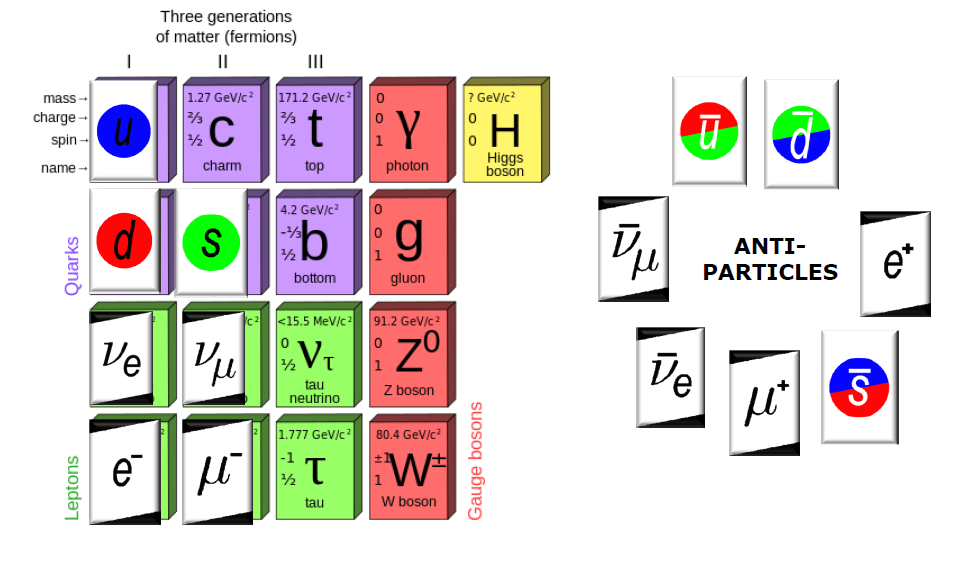}~\\*
\caption{
\fontfamily{lmss}\selectfont
Elementary particles of the Standard Model -- playfully, 
using the representations in the Quark Matter Card Game. 
Anti-particle representations are also included.
}
        \label{fig_3}
\end{figure}

Our current knowledge about the fundamental constituents of matter is
summarized in the so-called Standard Model of Particle Physics. The elementary
particles of the Standard Model can be arranged in the form of a 4x4 table,
where the first 3 columns represent the three generations or families of
matter-like particles (fermions) and the last column represents the
interaction-mediating particles (bosons).

A playful representation of the most frequent matter-like particles was worked
out in the year of  2011,
 in the form of so called Quark Matter Card Games, as illustrated on
Figure~\ref{fig_3}.  This representation is detailed in refs.~
\cite{Csorgo:2013xza,Csorgo:2013vza}.

In 2012,
as an extra bonus to this 4x4 table, the last missing piece, the so called
Higgs boson of the Standard Model was also discovered experimentally. 
A card game that popularizes the discovery of the Higgs boson is 
detailed elsewhere ~\cite{Csorgo:2013vza}.
Here we focus on the gamification and modelling of 
the properties of Quark Matter, the perfect fluid of quarks discovered at RHIC and confirmed at LHC.

The theory of the strong interactions, Quantum Chromo Dynamics (QCD) has some
mathematical properties that are analogous to the properties of the optical
colors. Due to this mathematical analogy the quarks can be modelled with cards
that have optical colors: quark cards are colored to red, green and blue, the
three fundamental colors  in the RGB color space.
One of the exact laws of  QCD is that only those
combinations of quarks are experimentally observable, that correspond to a
color neutral (white) combination of quarks.  One should also emphasize that
Color in Quantum Chromo Dynamics is not to be confused with the visible,
optical colors, but it can be understood as an optical model or analogy that
reflects well the mathematical properties of the physical theory QCD and that
analogy is used here to model strongly interacting fundamental particles called
quarks and anti-quarks.

For an introduction on the birth  of  the quark concept that lead the way to
the development of QCD as the theory of strong interactions as well as to the
first ideas on the analogy of optical colors to model certain symmetry
properties of the strong interactions, we recommend two early articles by Zweig
and Gell-Mann~\cite{Zweig:1981pd,GellMann:1981ph}.

\begin{figure}[ht]
        \centering
        \includegraphics[width=0.8\linewidth]{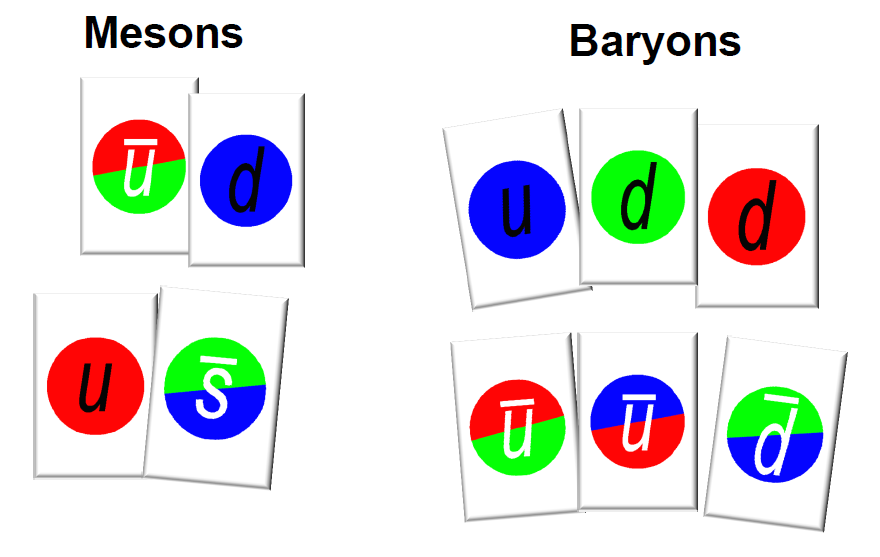}~\\*
        \caption{
\fontfamily{lmss}\selectfont
Mesons are colorless (white) combinations of a quark and an antiquark,
that is represented in the Quark Matter cards as a red, green or blue quark card 
matched with an anti-red (green/blue), anti-green (blue/red) or anti-blue (red/green)
pair of cards. Baryons are 
represented by a red, a green and a blue quark card, forming
also a colorless (white) combination of three quarks. 
Anti-baryons are also colorless, they can be formed from an anti-red, anti-green and anti-blue Quark Matter card.
}
        \label{fig_4}
\end{figure}

Importantly, another exactly satisfied law of elementary particle physics
states that for each particle, a there exists a corresponding anti-particle,
which is opposite in each properties to the given particle.  For example,
electron is an elementary particle with negative charge, so its antiparticle,
the positron has a positive charge.  But what is the opposite color to the red
color?  In the Quark Matter Card Games, we have chosen a combination of green
and blue colors to model (symbolically represent) the anti-red color, because
the green and blue combination supplements red to form a neutral, white color.
Similarly, anti-green is defined as a combination of blue and red, while
anti-blue is a red-green combination.  Three major groups of color neutral or
white particles can be formed: mesons or quark-antiquark color white bound
states, baryons (bound states of red, green and blue quarks) and anti-baryons
(bound states of anti-red, anti-green and anti-blue quarks), as indicated on
Figure~\ref{fig_4}.

\begin{figure}[h!t]
        \centering
        \includegraphics[width=0.7\linewidth,angle=-90]{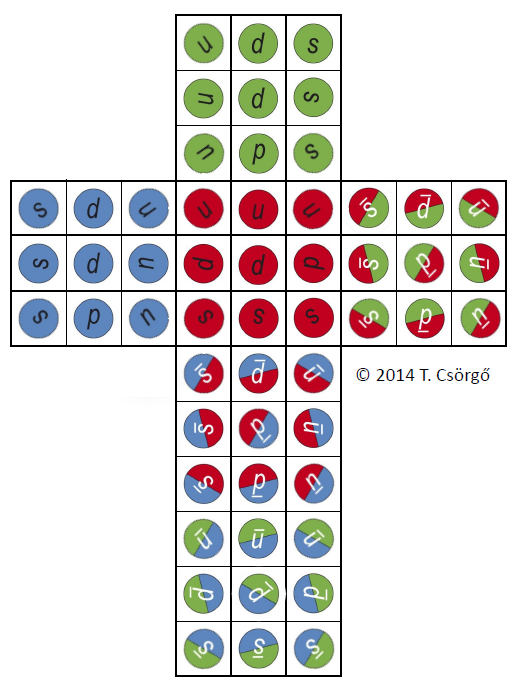}~\\*
        \caption{
\fontfamily{lmss}\selectfont
Layout of Qbe, the perfect Fluid of Quarks on Rubik's 3x3 Cube.
}
        \label{fig_5}
\end{figure}

With the help of the colored quarks and antiquarks, and the six faces of
Rubik's cube, 
one create a customized version of Rubik's cube in the following manner:
Three
faces that join in a single corner of the cube are selected to have red, green
and blue colors. The diagonally opposite corner of the cube is selected to be
the place where the anti-colored faces meet. The three most abundantly produced
quarks ($u$, $d$ and $s$) are also indicated on these little faces.  The
coloring scheme for the cube is such that quarks with a given color are on
opposite faces with anti-quarks with the corresponding anti-color. For example
the red quarks are opposite to the green/blue anti-quarks.  Thus the opposite
faces of Qbe combine to a white color, hence Qbe has an overall white
color. 
This  design, or the dressing up of  Rubik's cube as Qbe or
Quark Matter Cube is laid out on Figure ~\ref{fig_5}.

\begin{figure}[h!b]
        \centering
        \includegraphics[width=0.85\linewidth]{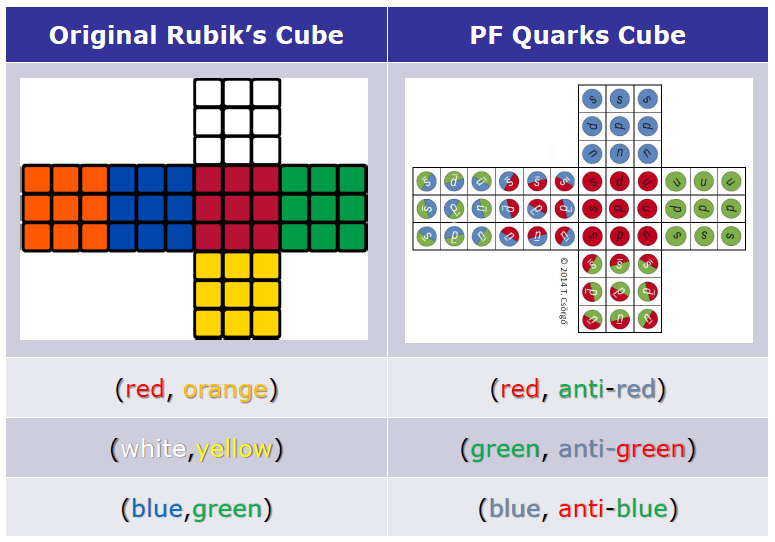}~\\*
        \caption{
\fontfamily{lmss}\selectfont
Comparison of the color scheme of Qbe with the color scheme of the original Rubik's 3x3 cube.
}
        \label{fig_6}
\end{figure}

Such a design can be well compared to the color scheme of the original Rubik's
cube.  This is illustrated on Figure~\ref{fig_6}. The Rubik design is dressing
up opposite faces with color and color + yellow color: the white face of
Rubik's cube is opposite to the yellow, red is opposite to orange and the blue
face is opposite to the green face. On Qbe, the three faces with the
 fundamental red, green and blue colored quarks  are placed opposite to the
three faces with the fundamental anti-colors: anti-red, anti-green and
anti-blue. This color scheme of Qbe reflects faithfully the overall color neutrality or whiteness of Quark Matter.

In the Early Universe, just a few microseconds after the Big Bang,
Quark Matter is created in a special way, namely the number of quarks and the number of anti-quarks were 
almost exactly the same at that time. This is property of the Early Universe
is faithfully represented: on Qbe the number of quarks 
is exactly the same as the number of antiquarks, as apparent from 
Figure ~\ref{fig_5}. In the deck of cards of the Quark Matter Card Game,
the number of quarks is larger than the number of anti-quarks, corresponding to the
properties of Quark Matter created in high energy heavy ion collisions at man-made accelerators.

\begin{figure}[h!t]
        \centering
        \includegraphics[width=1.0\linewidth]{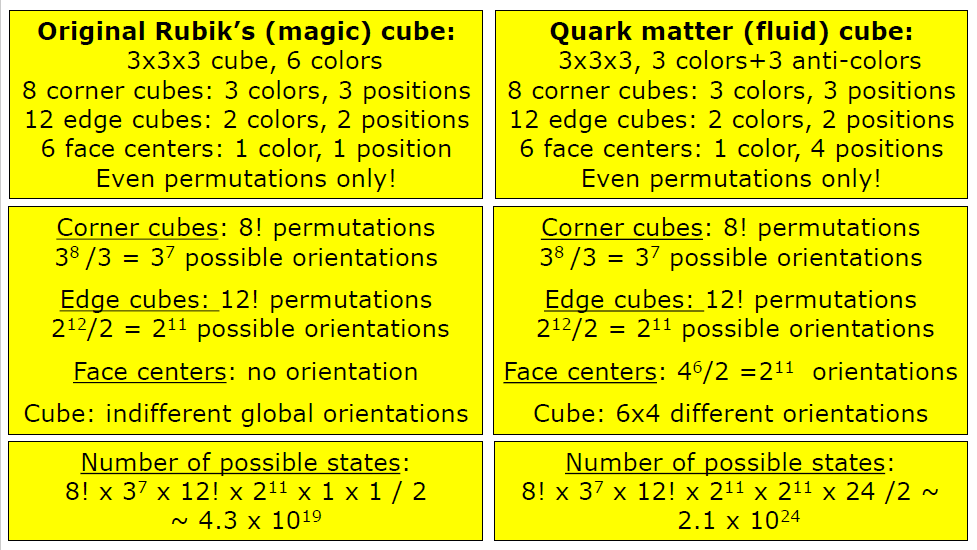}~\\*
        \caption{
\fontfamily{lmss}\selectfont
Comparison of mathematical properties of Qbe and Rubik's Cube.
}
        \label{fig_7}
\end{figure}

The mathematical properties, namely the possible number of color configurations
on Rubik's cube are compared to the properties of Qbe on
Figure~\ref{fig_7}.  We emphasize that the position of the $u$, $d$
and $s$ quarks in a heavy ion collision is physically a relevant quantity as the 
masses and other properties of these quarks vary. So we
suggest to distinguish the physical orientations of Qbe, which gives an extra
6x4 = 24 factor for its number of states. In addition, due to the  $u$,
$d$ and $s$ letters written on the facelets to represent quarks, the
face-center facelets are oriented so the total number of possible
configurations of Qbe, the Quark Matter cube is larger than the number of
states on Rubik's cube. The logarithm of the number of states corresponds to
the entropy content of these cubes. The entropy divided by volume defines their
entropy density.

\begin{figure}[h!b]
        \centering
        \includegraphics[width=0.95\linewidth]{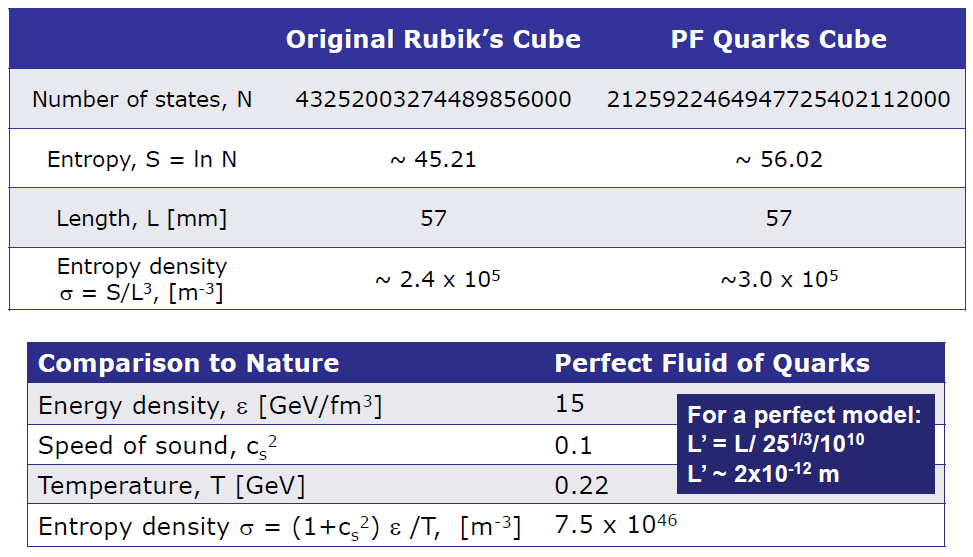}~\\*
        \caption{
\fontfamily{lmss}\selectfont
Comparison of physical properties of Qbe and Rubik's Cube. 
}
        \label{fig_8}
\end{figure}

The entropy density of Qbe the perfect fluid of quarks on Rubik's 3x3 cube can
be compared to the entropy density of quark matter created in heavy ion
collisions at RHIC and LHC accelerators. To have the same entropy density as
Quark Matter, Qbe should be scaled down too much, from 57 mm to 2x10$^{-12}$ m,
but instead of scaling the cube down, we suggest use Qbe as a model of Quark Matter that fits 
suitably the size of our hands.  The physical properties of Qbe the Quark Matter Cube can thus be
compared also to the physical properties of Quark Matter, as summarized on
Figure ~\ref{fig_8}.

\section{Discussion}

The connection between the symmetry properties of quarks and Rubik's cube with
twisted corner pieces has been noted by Golomb already in
1981~\cite{Golomb:1981ze}. His article determined the number of color
configurations on Rubik's cube as well.  Marx and collaborators considered
Rubik's cube as a kind of world model, with conservation laws and
transformation rules and they noted how baryons and mesons might be represented
with Rubik's cubes with twisted corner or edge cubelets~\cite{Marx:1982gg}.
Hofstadter picked up the idea of Golomb and noted the importance of variations
on the same theme as a key element to innovation. His 1982 article made
Rubik's Cube with a twisted or ``quarked" corner piece to the cover page of
Scientific American ~\cite{Hofstadter:1982dr}.  However, as far as we know,
Quark Matter with colored quarks and anti-quarks was not considered in the
context of a Rubik's cube before. 

Although the mathematical and engineering aspects of the Rubik's cube were
summarized already in 1987 by E. Rubik and collaborators~\cite{Rubik:1987aa},
some of the mathematical aspects of the Rubik's cube imposed deep and difficult
problems.  For example, the minimum number of rotations that are needed to
reach any given configuration from a perfectly ordered Rubik's cube (the so
called God's number) was proven to be 20 by Rokicki only in
2014~\cite{Rokicki:2014aa}. As far as I know the God's number for Qbe or other
generalized Rubik's cubes with oriented face centers is not yet determined.

The educational values of Rubik's cube in visual-spatial intelligence,
developing strategy, improving memorization, concentration and persistence
in problem solving as well as the marketing values of  Rubik's cube in  
popular Science, Technology, Engineering and Mathematics (STEM) were
overviewed recently in ref.  ~\cite{Kiss:2015mm}. 

Let us mention, that Rubik's cube was recently envisioned as a model for
describing the change of the interiors of black holes while emitting a Hawking
particle and thus decreasing the size and corresponding the entropy of a black
hole. This processs was conjectured to be analogous with solving the Rubik's
cube~\cite{Czech:2011wy}.  This analogy between an evaporating  black hole
to vacuum and solving the Rubik's cube from a large initial entropy / disorder
to a color ordered, zero entropy state may provide further inspiration for
follow-up  STEM gamification and outreach studies.

To illustrate that quite some  time and wisdom  might be needed to solve Qbe,
let us estimate how many rotations might be needed in every second, if 
we would try to solve it just by random rotations.  Our
Universe is about 13.8 $\times$ 10$^{9}$ years old and the number of states on
Qbe is given in Figure ~\ref{fig_7} as approximately  2.1$\times$ 10$^{24}$. As
the lifetime of our Universe converts to about  4.35 $\times$ 10$^{17}$
seconds, one would need to rotate the Qbe a littlebit more than 4.8 million
times in every second, for the entire lifetime of our Universe, to be able to
solve it just by random rotations.  Such a tremendous mindless effort can be
contrasted to the various records of solving Rubik's cube using skillful means
in speed cubing championships: The current world record for single time on a
3$\times$3$\times$3 Rubik's Cube was set by Feliks Zemdegs of Australia in 
December 2016 with a time of 4.73 seconds at the POPS Open 2016 competition in
Melbourne, Australia~\cite{CubeRecord:2017aa}.

\begin{figure}[h!t]
        \centering
        \includegraphics[width=1.0\linewidth]{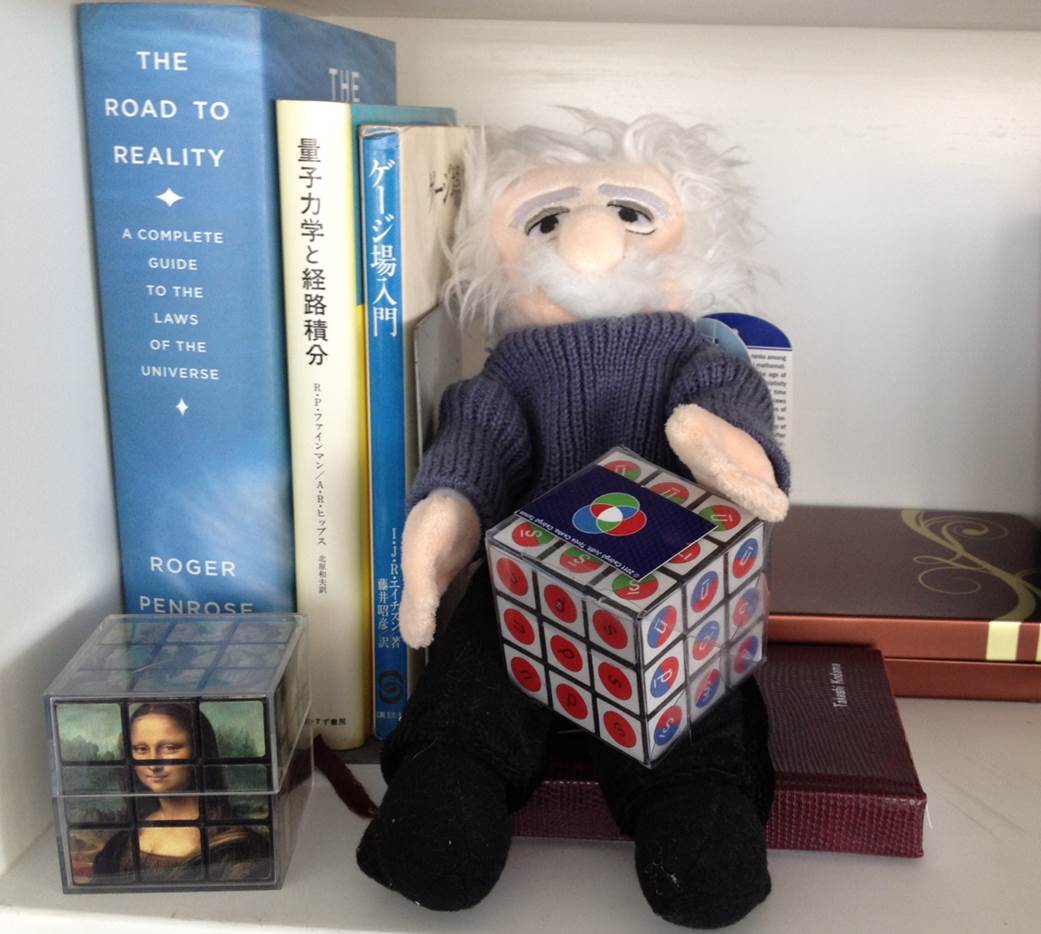}~\\*
        \caption{
\fontfamily{lmss}\selectfont
Figure of Albert Einstein, the smile of Mona Lisa and Qbe: Quark Matter 
on Rubik's 3x3 Cube, next to the Road to Reality: A Complete Guide to
the Laws of the Universe. Photo courtesy of prof. T. Kodama, Rio de Janeiro,
Brazil.
}

        \label{fig_9}
\end{figure}

Let us close this article by noting that  what we discussed here was just a toy
or a toy model, that does not have to be taken too seriously. In this sense
this outreach article is quite similar to many studies in science.  A model is
just a model, reflecting certain properties of the reality and is best
understood with a certain smiling playfulness, similar to the mysterious smile
on the face of Mona Lisa.  The Road to Reality is often a difficult one but our
journey may become much more enjoyable, perspiacious and lightsome if we
proceed with a touch of smiling wisdom, as illustrated on Figure ~\ref{fig_9}.

The Appendix of this contribution is organized as a handout booklet,
to be distributed with Qbes or Quark Matter Cubes.

\section*{Acknowledgments}

Inspiring discussions with S. Kiss are greatfully acknowledged.  The
presentation of Quark Matter on Rubik's 3x3 Cube at the WPCF 2014 conference in
Gy\"ongy\"os, Hungary has been partially supported by the OTKA NK 101438 grant.

\section*{Availability} 

Limited number of Qbes, Quark Matter on Rubik's Cubes were distributed first as
promotional gifts to the participants of the WPCF 2014 conference in 
Gy\"{o}ngy\"{o}s, Hungary. 
As science outreach gifts, they are also made available at the
BERA Shop in the Berkner Hall at Brookhaven National Laboratory, Upton, NY,
USA, since June 2015. A limited number Qbes is
available just as well in the Iijima Shoten in KEK, Tsukuba, Japan.

\vfill\eject

\vfill\eject

\section*{Legal info and image rights}

The ``Quark Matter Card Game" and the corresponding deck and representation of
elementary particles on cards, commercial and design rights and copyrights  are
reserved and rest with the Authors.  

The RUBIK'S CUBE in its three dimensional form and any graphic or photographic
representation of it, in any configuration, colored or uncolored, whether or
not it carries the RUBIK'S CUBE name or logo, is protected by intellectual
property laws throughout the world. Rubik's Brand Ltd. owns all the
international rights in the RUBIK trademarks and in the overall image of the
RUBIK'S CUBE.  The copyright belongs to Ern\H{o} Rubik, the originator of
RUBIK'S CUBE who has given Rubik's Brand Ltd. full and exclusive authority to
license and administer his rights, and to pursue by whatever legal means
necessary any infringement of such rights.

\vfill\eject

\section*{Handout for Qbe: Quark Matter on Rubik's Cube} 

{\bf Perfect Fluid Promotion:} 
Qbe or Quark Matter on Rubik's 3x3 Cube is not only fun but also a promotional
tool to introduce and illustrate certain unusual properties of the perfect
fluid of quarks. A perfect fluid can flow without internal dissipation.
So the perfect fluid of Quark Matter could be modelled faithfully 
by a perfectly lubricated Rubik's cube, suited perfectly for
speed-solving competitions.

{\bf Number of players:} Typically one person, but speed and memory cubing
competitions can be organized.

{\bf Object of the game:} 
The goal is to solve a fully scrambled Qbe - Quark Matter on Rubik's 3x3 Cube
by reaching its color-flavor-locked ground state. Flavor locking means in this
context that the letters representing d flavoured quarks on the face-center
facelets should point towards the corner of where the red, green and blue faces
meet and the letters $\overline{d}$ that stand for anti-d antiquarks on the
face-center facelets of the anti-colored faces should simultaneously be
pointing to the opposite corner of Qbe, where faces with anti-red, anti-green
and anti-blue colors meet.

{\bf The course of the game:}
The players inspect the thoroughly scrambled Qbe,
then place it back to the desk in front of them. They may use any of the agreed
methods (both hands, or in extreme cases, single hand, both feet, blindfolded,
underwater and so on) to solve Qbe. By rotating the sides of Qbe, they compete
to reach the desired (color ordered or color-flavor locked) ground state of Qbe.

Qbe, the Quark Matter on Rubik's 3x3 Cube is a three dimensional combination
puzzle that can be solved on beginner, intermediate or advanced levels:

1) {\it On  beginner level,} players do not know the how to solve the standard
Rubik's Cube. It is a challenging task to figure it out on your own, but it is
worth to try. Physicists or physics students are expected to be able to do the
first layer on their own and some may even be able to do the second one without
too much effort. Doing all the three layers on his own lasted several weeks
even for Mr. Rubik himself, but these days there are several public videos
that show how to solve the cube, see for example \linkurl{
https://www.youtube.com/watch?v=rmnSpUgOvyI}.  This way the players will be
able to solve the colors of Qbe. However, the orientation of the $d$-quarks
on the center facelets on each face may still point to random directions.

2) {\it On  intermediate level,} the goal is to reach the color-flavor locked
ground state. In this case, after the faces are color ordered, all the $d$
quarks in the centers should point to the corner where the red, green and blue
colors meet, and all the anti-$d$ quarks should point to the opposite corners,
where the faces with anti-red, the anti-green and anti-blue colors meet. This
means that the players have to change the orientation of the center pieces on
the faces of the cube without destroying the color order. This is also an
already solved problem, sometimes referred to as solving the Super-Cube,
custom-cube or picture-cube.  Without significant cubing experience, physicists
are not expected to figure this out on their own. To fix the direction of the
centers,  see e.g.  \linkurl{ https://www.youtube.com/watch?v=fk1eCZNCTB4 } .

3) {\it On an advanced level,} the players already know how to solve Qbe. But
they can still improve the time they need to do so, they can try to do this
blindfolded, by one hand, or may use any other of the several mind-boggling
methods that were developed recently for the emerging arts of speed and memory
cubing.

{\bf Recommended physics talking points} are listed as follows:

\begin{enumerate}

\item
{\bf Color:} Quark Matter is a colorless state, but locally colors are
free, deconfined, as most of the cubelets have a net color. Qbe is
decorated by colored quarks and anti-colored anti-quarks to illustrate a state
of matter called Quark Matter or Perfect Fluid of Quarks.  Quarks come in three
different colors: red, green and blue.  Antiquarks have anti-colors called
anti-red, anti-green and anti-blue, represented by the combination of
green/blue, blue/red and red/green colors, following the model developed for
the Quark Matter Card Games
~\cite{Csorgo:2012qm,Csorgo:2013xza,Csorgo:2013vza,Csorgo:2013vza}.  In the
ground state, the red face of Qbe is opposite to the anti-red, blue face is
opposite to anti-blue, green is opposite to anti-green. In a random state of
Qbe, locally the colors are not compensating each other to a color neutral,
white  or red-green-blue combination, however, adding all the colors on Qbe
results in an overall, globally white color, that models faithfully the
globally color white but locally colored property of the   Quark Matter state.

\item 
{\bf Flavor:} Quarks may have 6 different flavors, denoted as $u, d, s,
c, t$ and $b$.  On Qbe, only the first three flavors are utilized: $u,d$
and $s$.  These flavors correspond to the flavors of the most abundantly
produced quarks at RHIC and LHC.  Can you order the faces of Qbe by
the flavor?

\item
{\bf Baryon number:} The net baryon number of any system of quarks is defined as 
the number of quarks minus the number of anti-quarks, divided by 3. What is the 
net baryon number of Qbe in its ground (ordered) state? 
Do rotations (that mix the quarks and antiquarks of Qbe)  
modify its net baryon number?

\item {\bf Entropy density:} 
Quark Matter has a huge entropy density, $\sigma\approx 7.5 \times 10^{45}/\mbox{\rm m}^3$.
This can be 
compared to the huge number of physically
different states of Qbe. When the 24 possible orientation of  a
given cube in space as well as all the possible orientation of the center
pieces are also taken into account, the possible number of states 
of Qbe 
becomes a  huge number:
2,125,922,464,947,725,402,112,000 ($\approx 2.12\times 10^{24}$), 
a bit larger than Avogadro's number, $6.02 \times 10^{23}$.
Derive the entropy density of Qbe, given that an edge of Qbe is 57 mm.  

\item {\bf Perfect Fluidity:} 
A fluid is perfect if it has no internal dissipation. The resistence of a fluid
to internal friction/shearing motion is characterized by the so called
kinematic viscosity, denoted by $\eta/\sigma$.  This is somewhat analogous to
the resistance of the faces of Rubik's cube to rotation: in a perfect model of
a perfect fluid, a rotating outer third of the cube could keep on rotating
forever, without resistance. 
Due to dissipative forces, this rotation is coming to an end shortly on a
physical model like a Qbe.  Use this analogy to estimate the kinematic
viscosity $\eta/\sigma$ of Qbe, the Quark Matter on Rubik's 3x3 Cube,  if the
torque needed to rotate an outer third layer of Qbe is of the order of 0.1 Nm
and $\sigma$ is the entropy density of Qbe evaluated in item 4 above.  How far
Qbe is from the conjectured quantum limit for a perfect fluid,  $ \eta/\sigma =
\hbar /(4 \pi)$? 

\end{enumerate}

\end{document}